\documentstyle[12pt,epsf]{article}

\def\beq{\begin{equation}}
\def\eeq{\end{equation}}
\def\bea{\begin{eqnarray}}
\def\eea{\end{eqnarray}}

\begin{document}

\thispagestyle{empty}

\font\fortssbx=cmssbx10 scaled \magstep2
\hbox to \hsize{
\hbox{\fortssbx University of Wisconsin - Madison}
      \hfill$\vtop{
\hbox{\bf MADPH-99-1107}
\hbox{\bf AMES-HET-99-02}
\hbox{March 1999}}$ }

\vspace{.5in}

\begin{center}
{\large\bf Seasonal and energy dependence\\
of solar neutrino vacuum oscillations}\\
\vskip 0.4cm
{V. Barger$^1$ and K. Whisnant$^2$}
\\[.1cm]
$^1${\it Department of Physics, University of Wisconsin, Madison, WI
53706, USA}\\
$^2${\it Department of Physics and Astronomy, Iowa State University,
Ames, IA 50011, USA}\\
\end{center}

\vspace{.5in}

\begin{abstract}				    

We make a global vacuum neutrino oscillation analysis of solar neutrino
data, including the seasonal and energy dependence of the recent
Super--Kamiokande 708-day results. The best fit parameters for $\nu_e$
oscillations to an active neutrino are $\delta m^2 =
4.42\times10^{-10}$~eV$^2$, $\sin^22\theta = 0.93$. The allowed mixing
angle region is consistent with bi--maximal mixing of three neutrinos.
Oscillations to a sterile neutrino are disfavored. Allowing an enhanced
$hep$ neutrino flux does not significantly alter the oscillation
parameters.

\end{abstract}

\thispagestyle{empty}
\newpage

\section{Introduction}

For quite some time, measurements of solar neutrinos~\cite{solar,Kam}
have indicated a suppression compared to the expectations of the
standard solar model (SSM)~\cite{SSM}. This suppression may be explained
by assuming that $\nu_e$ from the Sun undergo vacuum oscillations as
they travel to the Earth~\cite{bpw81,justso}. Recent data from the
Super--Kamiokande (SuperK) experiment~\cite{SuperK708} also exhibit a
seasonal variation above that expected from the $1/(\rm distance)^2$
dependence of
the neutrino flux, which if verified would be a clear signal of vacuum
oscillations\cite{vlw,seasonal,bw98}.

In this Letter we determine the best fit vacuum oscillation parameters
to the combined solar neutrino data, including the 708--day SuperK
observations. For oscillations to an active neutrino there are subtle
changes in the allowed regions compared to fits with earlier
SuperK data~\cite{bw98,bks98}. Oscillations to a sterile neutrino are
disfavored. We examine the possibility that the solar $hep$
neutrino flux is enhanced compared to the SSM, and find that the best
fits are only marginally changed. We also find a solution with very
low $\delta m^2 \simeq 6\times10^{-12}$~eV$^2$ for oscillations to a
sterile neutrino or with an $hep$ enhancement.

\section{Fitting procedure}

The fitting procedure is described in detail in Ref.~\cite{bw98}. The
solar data used in the current fit are the $^{37}$Cl (1 data point) and
$^{71}$Ga (2 data points) capture rates~\cite{solar}, the latest SuperK
electron recoil energy spectrum with $E_e$ in the range 5.5 to
20~MeV~\cite{SuperK708} (18 data points), and the latest seasonal
variation data from SuperK for $E_e$ in the range 11.5 to
20~MeV~\cite{SuperK708} (8 data points). For $\nu_e$ oscillations to an
active neutrino ($\nu_\mu$ or $\nu_\tau$) we take into account the
neutral current interactions in the SuperK experiment.  We fold the
SuperK electron energy resolution~\cite{Kam} in the oscillation
predictions for the $E_e$ distribution. For all rates
that are annual averages we integrate over the variation in the
Earth-Sun distance. We do not include the SuperK day/night ratio, since
it is unity for vacuum oscillations.

To allow for uncertainty in the SSM prediction of the $^8$B neutrino
flux, we include as a free parameter $\beta$, the $^8$B neutrino flux
normalization relative to SSM prediction. In our fits with non--standard
$hep$ neutrinos we also include an arbitrary $hep$ normalization
constant, $\gamma$. For the SSM predictions we adopt the results of
Ref.~\cite{SSM}.

\section{Solutions with oscillations to an active neutrino}

Figure~1a shows the 95\%~C.L. allowed regions for the combined
radiochemical ($^{37}$Cl and $^{71}$Ga) capture rates and the SuperK
electron recoil spectrum, including its normalization. For the
95\%~C.L. region in a three--parameter fit ($\delta m^2, \sin^22\theta,
\beta$) we include all solutions with $\chi^2 < \chi^2_{min} + 8$. On
the boundary curves of the fit to the radiochemical data, the faster
oscillations are due to the lower energy neutrinos (mainly $pp$ and
$^7$Be), and the slower oscillations are due to the $^8$B neutrinos. The
five regions allowed by the SuperK spectrum data correspond roughly to
having a mean Earth--Sun distance equal to (in increasing order of
$\delta m^2$) ${1\over2}$, ${3\over2}$, ${5\over2}$, ${7\over2}$, or
${9\over2}$ oscillation wavelengths for a typical $^8$B neutrino energy.
Following the notation of Ref.~\cite{bw98}, we label these regions A, B,
C, D, and E, in order of ascending $\delta m^2$.

In Figure~1b we show the allowed regions for the combined radiochemical
and SuperK spectrum data (the solid curve).  Also shown in Fig.~1b is
the region excluded by the seasonal SuperK data at 68\%~C.L. (we show
the 68\%~C.L. region since almost none of the parameter space is
excluded by the seasonal variation at 95\%~C.L.).

Finally, in Fig.~1c we show the 95\%~C.L. allowed regions obtained with
all of the data (radiochemical, SuperK spectrum, and SuperK seasonal).
Only regions A, C, and D are allowed at 95\%~C.L. by the combined data
set. The best fits in each of the subregions are shown in
Table~\ref{active}. Region B, which was allowed in previous
fits~\cite{bw98,bks98}, is now excluded. The overall best fit parameters
are in region C
\beq
\delta m^2 = 4.42\times10^{-10}{\rm~eV}^2 \,, \qquad
\sin^22\theta = 0.93 \,, \qquad
\beta = 0.78 \,,
\label{bestfit}
\eeq
with $\chi^2/DOF = 33.8/26$, which corresponds to a goodness--of--fit of
14\% (the goodness--of--fit is the probability that a random repeat of
the given experiment would observe a greater $\chi^2$, assuming the model
is correct). The next best fit is in region $D$ with
\beq
\delta m^2 = 6.44\times10^{-10}{\rm~eV}^2 \,, \qquad
\sin^22\theta = 1.00 \,, \qquad
\beta = 0.80 \,,
\label{bestfit2}
\eeq
with $\chi^2/DOF = 36.7/26$, which corresponds to a goodness--of--fit of
8\%. Finally, region A is also allowed, with best fit
\beq
\delta m^2 = 6.5\times10^{-11}{\rm~eV}^2 \,, \qquad
\sin^22\theta = 0.70 \,, \qquad
\beta = 0.94 \,,
\label{bestfit3}
\eeq
with $\chi^2/DOF = 38.4/26$, which corresponds to a goodness--of--fit of
6\%. Regions C and D are consistent with bi--maximal\cite{bimax} or
nearly bi--maximal\cite{maxosc} three--neutrino mixing models that can
describe both solar and atmospheric neutrino data. Regions B and E are
excluded at 95\%~C.L.

We see from Table~\ref{active} that generally the higher $\delta m^2$
solutions fit the $E_e$ spectrum better. This is evident in Fig.~2,
which shows the $E_e$ spectrum predictions for the solutions in
Table~\ref{active}, and the latest SuperK data~\cite{SuperK708}. The
higher $\delta m^2$ solutions do worse for the radiochemical
experiments, primarily because the seasonal variation in the oscillation
argument is larger in these cases for the $^7$Be neutrinos,
causing more smearing of the
oscillation probability. The annual--averaged suppression of $^{7}$Be
neutrinos is about 93\% for solution A, 66\% for solutions B and C, and
57\% for solutions D and E. The best fit in each region generally lies
near a local minimum for the $^7$Be neutrino contribution, as might be
expected from model independent analyses without oscillations that
indicate suppression of $^7$Be contributions~\cite{modelindep}.

Results for the seasonal variation of the solutions in
Table~\ref{active} are shown in Fig.~3, along with the latest SuperK
data~\cite{SuperK708}. Also shown is the best fit for no oscillations
with arbitrary $^8$B neutrino normalization; the latter has a small
seasonal dependence due to the variation of the Earth-Sun
distance. Current seasonal data do not provide a strong constraint, but
clearly solutions A, C, and D fit the data better than the
no--oscillation curve.

Also included in Table~\ref{active} is a solution with very low $\delta
m^2$, which we call solution Z, that was pointed out in
Ref.~\cite{kp96}. It is excluded at 95\%~C.L., but marginally allowed at
99\%~C.L. (as is solution B). It corresponds roughly to a mean
Earth--Sun distance equal to ${1\over2}$ of the oscillation wavelength
(maximal supression) for $^7$Be neutrinos. Although it does very well in
suppressing the $^7$Be neutrinos (by about 94\% after accounting for
seasonal averaging), it does not show a significant seasonal effect
beyond that provided by the variation of the Earth-Sun distance (see
Fig.~3b).

\section{Solutions with oscillations to a sterile neutrino}

A similar analysis may be made with solar $\nu_e$ oscillating into a
sterile neutrino species. Since the sterile neutrino does not interact
in any of the detectors, it is harder to reconcile the differing rates
of the $^{37}$Cl and SuperK experiments~\cite{vlw}. The best fit
parameters in each of the regions are shown in Table~\ref{sterile}.  For
sterile neutrinos the overall best fit parameter is again in region C
with $\chi^2/DOF = 42.4/26$, which is excluded at 97.7\%~C.L. Therefore
oscillations to sterile neutrinos are highly disfavored. Interestingly,
the fit for solution Z is comparable to the other best--fit solutions in
the sterile case because its strong suppression of $^7$Be neutrinos
helps account for the difference between the $^{37}$Cl and SuperK rates
(note the $\chi^2$ values for the $^{37}$Cl data in
Table~\ref{sterile}).

\section{Solutions with non-standard contributions from $hep$ neutrinos}

Recently it has been speculated that the rise in the SuperK $E_e$
spectrum at higher energies could be due to a larger than expected $hep$
neutrino flux contribution~\cite{frere,bk98}. While the maximum energy
of the $^8$B neutrinos is about 15~MeV, the $hep$ neutrinos have maximum
energy of 18.8~MeV. In the SSM the total flux of $^8$B neutrinos is
about 2000 times that of the $hep$ neutrinos~\cite{SSM}, and the $hep$
contribution to the SuperK experiment is negligible. However, there is a
large uncertainty in the low energy cross section for the reaction $^3He
+ p \rightarrow ^4He + e^+ + \nu_e$ in which the $hep$ neutrinos are
produced, and an $hep$ flux much larger than the SSM value may not be
unreasonable~\cite{frere,bk98}. A large enhancement of the $hep$ contribution
could in principle account for the rise in the $E_e$ spectrum at higher
energies seen by SuperK. The $hep$ flux normalization $\gamma$ can be
determined once there are sufficient events in the region $E_e >
15$~MeV. SuperK measurements of the $E_e$ spectrum in the range
17--25~MeV already place the upper limit $\gamma < 8$ at
90\%~C.L.~\cite{SuperK708}, assuming no oscillations.

It should first be noted that although an enhanced $hep$ contribution
{\it without} neutrino oscillations can provide a good fit to the SuperK
data for $5.5 < E_e < 14$~MeV, it cannot also account for the $^{37}$Cl
and $^{71}$Ga rates even with arbitrary $^8$B, $hep$, {\it and} $^7$Be
neutrino flux normalizations. The overall best fit in this case occurs
with no $^7$Be contribution, and has $\chi^2/DOF = 53.7/26$, which is
excluded at 99.9\%~C.L. The contributions of the $pp$ neutrinos, plus
the reduced amount of $^8$B neutrinos needed to explain SuperK, give a
rate for the radiochemical experiments that is too high, even when the
$^7$Be contribution is ignored.

One can also ask what happens if an enhanced $hep$ contribution is
combined with neutrino oscillations~\cite{bk98}, although the motivation
for enhancing the $hep$ neutrino flux is not strong here since vacuum
oscillations already can explain the rise in the SuperK $E_e$ spectrum.
Table~\ref{hep} shows the best fits in each region when an arbitrary
$\gamma$ is allowed. The overall best fit parameters in this case are
again in region C with $\chi^2/DOF = 32.8/25$, which corresponds to a
goodness--of--fit of 14\%. The fits for most of the regions are not
significantly improved from the standard $hep$ flux case, and the fitted
oscillation parameters are little changed. The only exceptions are
solutions E and Z, which originally could not explain the SuperK
spectrum as well, but with the addition of extra $hep$ neutrinos can now
also provide a reasonable fit to all of the data. Regions A, C, D, E,
and Z are all allowed at 95\%~C.L. when an $hep$ enhancement is
included. However, the preferred values of $\gamma$ exceed the current
bound from SuperK, so that the role of an $hep$ flux enhancement
appears to be minimal.

\section{Summary and discussion}

The latest SuperK solar neutrino data suggest there is a seasonal
variation in the solar neutrino flux. The hypothesis that solar $\nu_e$
undergo vacuum oscillations to an active neutrino species provides a
consistent explanation of all the solar data, with a best fit given by
oscillation parameters $\delta m^2 = 4.42\times10^{-10}$~eV$^2$ and
$\sin^22\theta = 0.93$. Oscillations to sterile neutrinos are ruled out
at 97.7\%~C.L., and fits with an enhanced $hep$ neutrino flux do not
significantly alter the fit results.

The existence of vacuum neutrino oscillations can be confirmed with more
data from SuperK on the seasonal variation of the $^8$B neutrino
flux. The spectrum and seasonal variations of $^8$B neutrinos can also
be measured in the Sudbury Neutrino Observatory (SNO)~\cite{SNO} and
ICARUS~\cite{ICARUS} experiments. The line spectrum of the $^7$Be
neutrinos gives larger seasonal variations than $^8$B~\cite{vlw,seasonal}
and these may
be observable with increased statistics in $^{71}$Ga experiments, or in
the BOREXINO experiment\cite{borexino}, for which $^7$Be neutrinos
provide the dominant signal. Accurate measurements of the seasonal
variation in these experiments should be able to distinguish between the
different vacuum oscillation scenarios~\cite{bw98}, providing a unique
solution to the solar neutrino problem.

\section*{Acknowledgements}

We thank S. Pakvasa for a stimulating discussion and B. Balantekin for
useful conversations. This work was supported in part by the
U.S. Department of Energy, Division of High Energy Physics, under Grants
No.~DE-FG02-94ER40817 and No.~DE-FG02-95ER40896, and in part by the
University of Wisconsin Research Committee with funds granted by the
Wisconsin Alumni Research Foundation.


\newpage


\begin{table}
\caption[]{Best fit parameters and $\chi^2$ for various vacuum
oscillation solutions with $\nu_e$ oscillations to an active neutrino.}
\label{active}
\vspace{0.5 cm}

\footnotesize\tabcolsep=.4em
\centering\leavevmode
\begin{tabular}{|c|ccc|cccc|c|}
\hline
& \multicolumn{3}{c|}{Parameters}   &   &    & $\chi^2$ & & $\chi^2$ \\
Solution & $\delta m^2$  & $\sin^22\theta$ & $\beta$
& $^{37}$Cl & $^{71}$Ga & spectrum & seasonal & total \\ 
& ($10^{-10}$~eV$^2$) &&&&&&& \\ \hline
A & 0.65 & 0.70 & 0.95 & 4.7 & 1.2 & 25.8 &  6.7 & 38.4 \\ \hline
B & 2.49 & 0.80 & 0.79 & 8.9 & 1.1 & 30.4 &  5.4 & 45.8 \\ \hline
C & 4.42 & 0.93 & 0.78 & 6.7 & 4.4 & 17.4 &  5.3 & 33.8 \\ \hline
D & 6.44 & 1.00 & 0.80 & 7.5 & 5.2 & 17.6 &  6.4 & 36.7 \\ \hline
E & 8.61 & 1.00 & 0.81 & 7.7 & 5.3 & 20.4 &  8.7 & 42.1 \\ \hline
Z & 0.06 & 1.00 & 0.47 & 5.0 & 2.5 & 24.8 & 13.3 & 45.6 \\ \hline
\end{tabular}

\end{table}


\begin{table}
\caption[]{Best fit parameters and $\chi^2$ for various vacuum
oscillation solutions with $\nu_e$ oscillations to a sterile neutrino.}
\label{sterile}
\vspace{0.5 cm}

\footnotesize\tabcolsep=.4em
\centering\leavevmode
\begin{tabular}{|c|ccc|cccc|c|}
\hline
& \multicolumn{3}{c|}{Parameters}    &    &    & $\chi^2$ & & $\chi^2$ \\
Solution & $\delta m^2$  & $\sin^22\theta$ & $\beta$
& $^{37}$Cl & $^{71}$Ga & spectrum & seasonal & total \\ 
& ($10^{-10}$~eV$^2$)&&&&&&& \\ \hline
A & 0.64 & 0.62 & 0.96 & 18.9 & 1.4 & 29.3 &  6.7 & 56.3 \\ \hline
B & 2.49 & 0.73 & 0.82 & 21.2 & 1.4 & 33.1 &  5.3 & 61.0 \\ \hline
C & 4.42 & 0.89 & 0.84 & 15.8 & 2.7 & 18.5 &  5.4 & 42.4 \\ \hline
D & 6.42 & 0.97 & 0.88 & 15.3 & 5.2 & 17.3 &  5.9 & 43.7 \\ \hline
E & 8.62 & 1.00 & 0.91 & 15.7 & 5.5 & 19.2 &  8.1 & 48.5 \\ \hline
Z & 0.06 & 1.00 & 0.48 &  6.2 & 2.4 & 25.5 & 11.5 & 45.6 \\ \hline
\end{tabular}

\end{table}


\begin{table}
\caption[]{Best fit parameters and $\chi^2$ for various vacuum
oscillation solutions with $\nu_e$ oscillations to an active neutrino
and a non--standard $hep$ contribution $\gamma$ times the SSM prediction.}
\label{hep}
\vspace{0.5 cm}

\footnotesize\tabcolsep=.4em
\centering\leavevmode
\begin{tabular}{|c|cccc|cccc|c|}
\hline
& \multicolumn{4}{c|}{Parameters}    &     &  & $\chi^2$ & & $\chi^2$ \\
Solution & $\delta m^2$  & $\sin^22\theta$ & $\beta$ & $\gamma$
& $^{37}$Cl & $^{71}$Ga & spectrum & seasonal & total \\ 
& ($10^{-10}$~eV$^2$)&&&&&&&& \\ \hline
A & 0.65 & 0.70 & 0.94 &  7 & 4.4 & 1.3 & 26.0 &  6.7 & 38.3 \\ \hline
B & 2.50 & 0.76 & 0.74 & 38 & 8.5 & 1.1 & 27.9 &  5.8 & 43.2 \\ \hline
C & 4.42 & 0.90 & 0.74 & 35 & 6.2 & 3.4 & 17.6 &  5.6 & 32.8 \\ \hline
D & 6.44 & 0.98 & 0.76 & 50 & 6.2 & 4.5 & 17.4 &  7.1 & 33.2 \\ \hline
E & 8.61 & 0.97 & 0.75 & 66 & 5.9 & 4.3 & 18.4 & 10.3 & 38.9 \\ \hline
Z & 0.06 & 0.98 & 0.44 & 52 & 2.7 & 2.2 & 19.3 & 14.9 & 39.1 \\ \hline
\end{tabular}

\end{table}

\clearpage


\begin{figure}
\centering\leavevmode
\epsfxsize=6in\epsffile{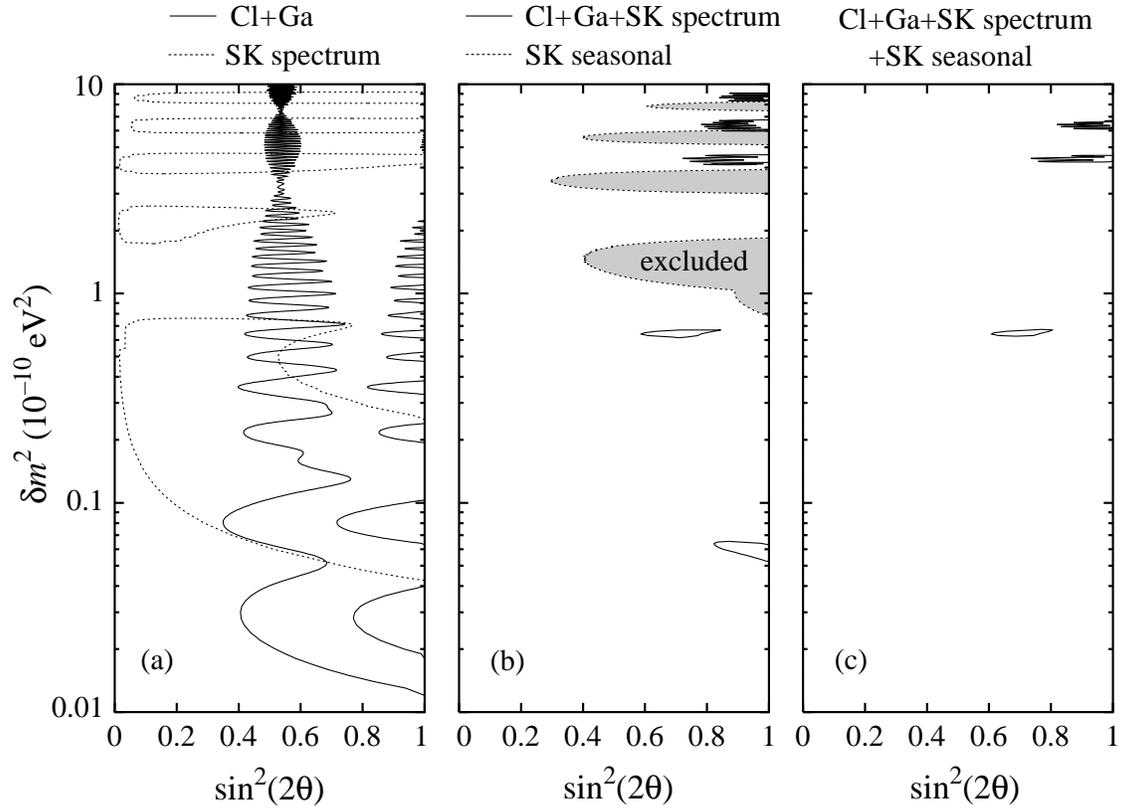}

\bigskip
\caption[]{\label{fig1} Regions in $\sin^22\theta$--$\delta m^2$
parameter space (a) allowed at 95\%~C.L. by the $^{37}$Cl and $^{71}$Ga
capture rates (solid curve) and the Super--Kamiokande $E_e$ spectrum
(dotted), (b) allowed at 95\%~C.L. by the combined $^{37}$Cl and
$^{71}$Ga capture rates and Super--Kamiokande $E_e$ spectrum (solid)
and excluded at 68\%~C.L. by the Super--Kamiokande seasonal data
(shaded), and (c) allowed at 95\%~C.L. by all the data.}
\end{figure}


\begin{figure}
\centering\leavevmode
\epsfxsize=4in\epsffile{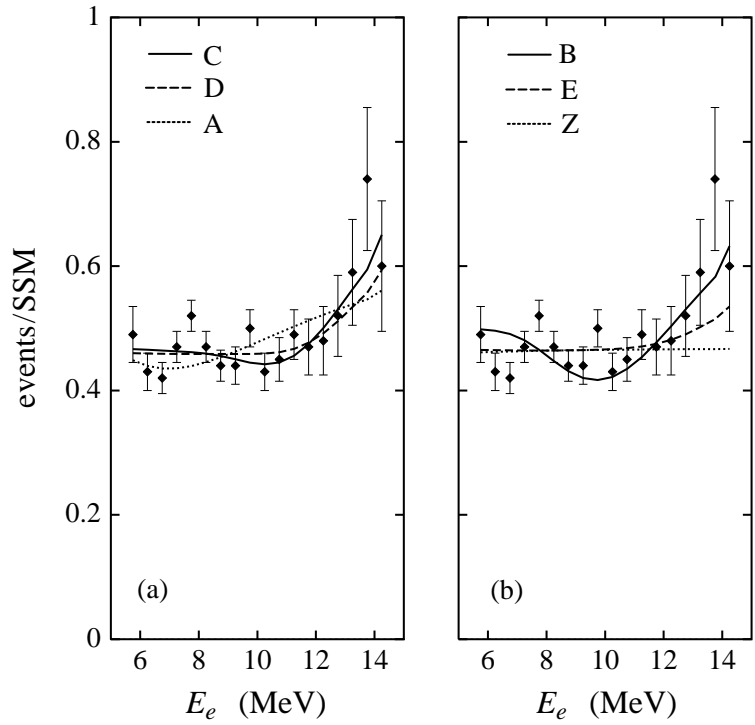}

\bigskip
\caption[]{\label{fig2} Ratio of predicted $E_e$ spectrum in
Super--Kamiokande to the SSM for (a) models favored by a global fit,
C (solid curve), D (dashed), and A (dotted), and (b) models disfavored
by a global fit, B (solid), E (dashed) and Z (dotted).}
\end{figure}


\begin{figure}
\centering\leavevmode
\epsfxsize=4in\epsffile{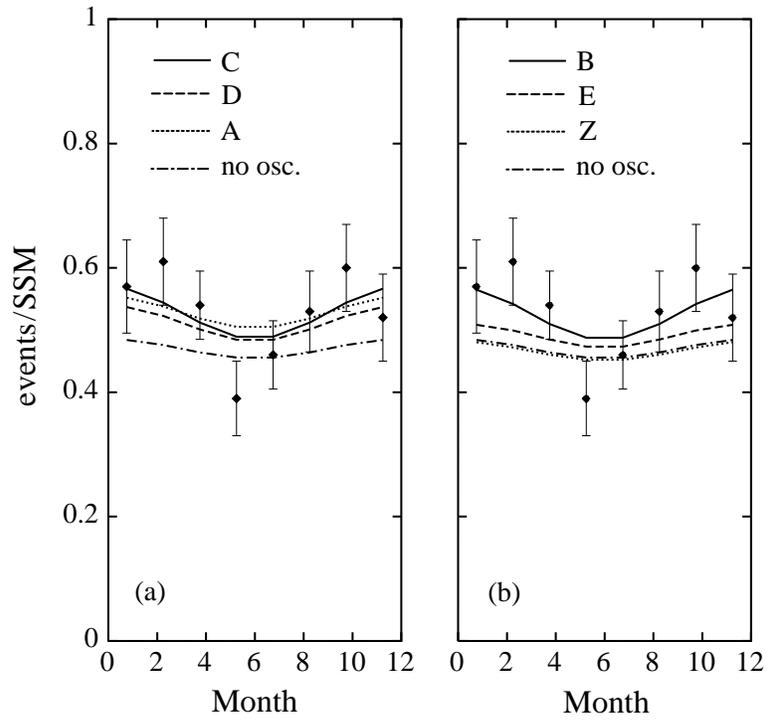}

\bigskip
\caption[]{\label{fig3} Ratio of predicted rate to SSM prediction versus
time of year in Super--Kamiokande for (a) models favored by a global fit,
C (solid curve), D (dashed), and A (dotted),  and (b) models
disfavored a by global fit, B (solid), E(dashed), and Z (dashed). Results for
no oscillations with $^8$B normalization of 0.47 are shown by the dash-dotted
curves.}
\end{figure}

\end{document}